\documentclass{aastex}\usepackage{emulateapj5,mathptmx}

\newcommand{\fun}{\hbox{\ erg cm$^{-2}$ s$^{-1}$} }
\newcommand{\lun}{\hbox{\ erg s$^{-1}$} }
\newcommand{\minus}{\hbox{--}}
\def\Chandra{{\it Chandra~}}
\def\XMM{{\it XMM-Newton~}}

\begin{document}
\submitted{To appear on the Astronomical Journal (Jan 2004)}

\title{Chandra and XMM-Newton Observations of RDCS1252.9\minus2927, \\
A Massive Cluster at \, \lowercase{$z=1.24$} \altaffilmark{1}}
\altaffiltext{1}{Based in part on observations
 obtained at the European Southern Observatory using the ESO Very
 Large Telescope on Cerro Paranal (ESO program 166.A-0701).}

\author{P. Rosati\altaffilmark{2}, P. Tozzi\altaffilmark{3},
S. Ettori\altaffilmark{2}, V. Mainieri\altaffilmark{2,12},
R. Demarco\altaffilmark{2,11}, S. A. Stanford\altaffilmark{4,5},
C. Lidman\altaffilmark{2}, M. Nonino\altaffilmark{3},
S. Borgani\altaffilmark{6}, R. Della Ceca\altaffilmark{7},
P. Eisenhardt\altaffilmark{8}, B.P. Holden\altaffilmark{9},
C. Norman\altaffilmark{10}
}

\affil{$^2$European Southern Observatory, Karl-Schwarzschild-Strasse
2, D-85748 Garching, Germany} 
\affil{$^3$INAF, Osservatorio
Astronomico di Trieste, via G.B. Tiepolo 11, I--34131, Trieste, Italy}
\affil{$^4$Department of Physics, University of California, Davis, CA 95616}
\affil{$^5$Institute of Geophysics and Planetary Physics, LLNL,
Livermore, CA 94551} 
\affil{$^6$Dip. di
Astronomia dell'Universit\`a, via G.B. Tiepolo 11, I--34131, Trieste,
Italy} 
\affil{$^7$INAF, Osservatorio Astronomico di Brera, 
via Brera 28, I--20121, Milano, Italy} 
\affil{$^8$Jet
Propulsion Laboratory, Mail Stop 169-327, California Institute of
Technology, Pasadena, CA 91109}
\affil{$^9$Lick Observatory, University of California, Santa Cruz, CA 95064} 
\affil{$^{10}$Department of Physics and Astronomy, Johns Hopkins
University, Baltimore, MD 21218} 
\affil{$^{11}$Institut d'Astrophysique de Paris, 98bis Boulevard Arago, F-75014 Paris, France}
\affil{$^{12}$Dip. di Fisica, Universit\`a degli Studi di Roma Tre, 
I-00146 Roma, Italy}

\begin{abstract}

We present deep \Chandra and \XMM obervations of the galaxy cluster
RDCS1252.9\minus2927, which was selected from the ROSAT Deep Cluster
Survey (RDCS) and confirmed by extensive spectroscopy with the VLT at
redshift $z=1.237$. With the \Chandra data, the X-ray emission from
the intra-cluster medium is well resolved and traced out to 500 kpc,
thus allowing a measurement of the physical properties of the gas with
unprecedented accuracy at this redshift. We detect a clear 6.7 keV
Iron K line in the \Chandra spectrum providing a redshift within 1\%
of the spectroscopic one. 
By augmenting our spectroscopic analysis
with the \XMM data (MOS detectors only), we significantly narrow down
the $1\sigma$ error bar to 10\% for the temperature and 30\% for the
metallicity, with best fit values $kT = 6.0^{+0.7}_{-0.5}$ keV, $Z =
0.36^{+0.12}_{-0.10} \, Z_\odot$.  In the likely hypothesis of
hydrostatic equilibrium, we measure a total mass of $M_{500} = (1.9
\pm 0.3) 10^{14} h_{70}^{-1} M_{\odot}$ within $R_{\Delta=500}\simeq
536$ kpc.  Overall, these observations imply that RDCS1252.9\minus2927
is the most X-ray luminous and likely the most massive bona-fide
cluster discovered to date at $z>1$. When combined with current
samples of distant clusters, these data lend further support to a mild
evolution of the cluster scaling relations, as well the metallicity of
the intra-cluster gas. Inspection of the cluster mass function in
the current cosmological concordance model
$(h,\Omega_m,\Omega_\Lambda)=(0.7,0.3,0.7)$ and $\sigma_8=0.7-0.8$ shows
that RDCS1252.9\minus2927 is an $M^\ast$ cluster at $z=1.24$, in
keeping with number density expectations in the RDCS survey volume.

\end{abstract}

\keywords{X-rays: galaxies: clusters --- galaxies: clusters: individual 
(RDCS 1252.9-2927) --- cosmology: observations}

\newpage

\section{INTRODUCTION}

X-ray studies of galaxy clusters over the last decade have driven
considerable observational progress in tracing the evolution of their
global physical properties. Based on X-ray selected samples covering a
wide redshift range, convincing evidence has emerged for modest
evolution of both the space density of the bulk of X-ray
clusters and their thermodynamical properties since $z\approx 1$ (see
Rosati et al. 2002 for a review).  With the advent of \Chandra and \XMM,
and their unprecedented sensitivity and angular resolution, these
studies have been extended beyond redshift unity and, in low redshift
clusters, have revealed the complexity of the thermodynamical
structure of the Intra-Cluster Medium (ICM) (e.g. Fabian et
al. 2003b).  Specifically, deep \Chandra observations of the handful
of clusters known to date at $z>1$ (Stanford et al. 2001, 2002) have
shown, for the first time at such large look-back times, the structure
of the ICM at scales below 100 kpc and have allowed emission weighted
temperatures to be measured.  The new \Chandra data have provided a
crude determination of cluster scaling relations at large lookback
times and a first study of their evolutionary trends (e.g. Holden et
al. 2002, Vikhlinin et al. 2002, Ettori et al. 2003b). In more distant
and complex systems, such as putative proto-clusters dominated by a powerful
radio galaxy, \Chandra deep pointings have only revealed non-thermal
components in the diffuse plasma so far (Fabian et al. 2003a, Scharf
et al. 2003). \XMM observations of very distant clusters, although
affected by source confusion in some circumstances, have the ability
to collect a large number of photons, thus improving temperature
determinations and allowing an estimate of the ICM metallicity
(Hashimoto et al. 2002, Tozzi et al. 2003).

The discovery and the study of systems beyond redshift unity provides the
strongest leverage for testing cluster formation scenarios. This,
however, has been a challenging task with current X-ray searches due to
the limited survey areas covered at faint fluxes. In this paper, we
present \Chandra and \XMM observations of the fourth cluster at $z>1$
discovered in the {\it ROSAT} Deep Cluster Survey (RDCS, Rosati et
al. 1998) at the very limit of the {\it ROSAT} sensitivity:
RDCS1252.9\minus2927 which has been confirmed at $z=1.237$ with an
extensive spectroscopic campaign carried out with the VLT.  \Chandra
observations of the other three distant RDCS clusters, RDCS0910+5422
($z=1.10$), RDCS0848.9+4452 ($z=1.265$), RDCS0848.6+4453 ($z=1.273$)
were presented in Stanford et al. 2002, 2001.  We describe the optical
and near-infrared data for RDCS1252 elsewhere (Rosati et al. in
preparation, Lidman et al. 2003), while we focus here on the X-ray
observations carried out with \Chandra, augmented with a partial \XMM
data set.  We derive physical parameters of the ICM and measure gas
metallicity from the clear presence of the iron K line in the X-ray
spectrum. We also derive the total mass of the cluster with a 16\%
accuracy by resolving the gas profile with the \Chandra data and by
combining \Chandra and \XMM spectra to improve the temperature
determination.  Our analysis takes advantage of the complementarity
between the XMM and \Chandra data sets: the superb angular resolution of
\Chandra is used to study morphological features of the ICM and to
flag point sources contaminating the cluster emission, while the XMM
data are used to boost the signal-to-noise of the diffuse component
thus improving its spectral analysis.  Overall, these data imply that
RDCS1252.9\minus2927 is the most X-ray luminous and likely the most
massive bona-fide cluster discovered to date at $z>1$.

$H_0=70\ {\rm km}\ {\rm s}^{-1}\,{\rm Mpc}^{-1}, \,\Omega_m=0.3, \,
\Omega_\Lambda=0.7$ are adopted throughout this paper.

\section{OBSERVATIONS AND X-RAY DATA REDUCTION}

\subsection{Discovery of RDCS1252.9\minus2927}

 RDCS1252.9\minus2927 (hereafter RDCS1252 for brevity) was selected as
 an extended X-ray source (with a significance of $3.2\sigma$) in the
 RDCS, which used a wavelet-based algorithm to detect and characterize
 X-ray sources in 180 ROSAT/PSPC archival fields down to $f_{\rm
 lim}$(0.5-2 keV)$=10^{-14}\fun$ (Rosati et al. 1998).  The source was
 found in the field with ROSAT ID WP300093 (exposure time $=15.7$
 ksec) at an off-axis angle of 13.9\arcmin\ with 31 net counts,
 corresponding to a flux of $(2.5\pm0.9) 10^{-14}\fun$ in the 0.5-2 keV
 band. At fluxes $\sim\!  2\times 10^{-14}\fun$, the RDCS covers an
 effective area of 5 deg$^2$ and has maximum sensitivity for $L^\ast$
 clusters at $z\gtrsim 1$\footnote{ We note that observations
 suggest $L_X^\ast(z=1)\simeq 10^{44}\lun$, in the 0.5-2 keV band
 (Rosati et al. 2002).}.  As a result, the four RDCS clusters at $z>1$,
 with $L_X\lesssim L_X^\ast$, were found in the faintest flux bin
 (Rosati et al. 1999, Stanford et al. 02).  A 30 minute $I$-band image
 obtained at the CTIO 4-m telescope with the Prime Focus camera in
 February 1997, revealed only a faint ($I\simeq 21.7$) galaxy pair
 very close to the X-ray centroid position. As part of a program to
 follow-up faint RDCS cluster candidates in the near-IR, $J$ and $K$
 band imaging was obtained with the SOFI camera at the NTT in November
 1998, which showed a clear overdensity of red galaxies with
 $J-K\simeq 1.9$, typical of early type galaxies at $z>1$ (see Lidman
 et al. 2003).  RDCS1252 has more recently been the core of a VLT
 Large Programme which included optical imaging with FORS2, deep near
 IR imaging with ISAAC (Lidman et al. 2003) and extensive spectroscopy
 with FORS2. Results from this program are described elsewhere; to
 date 36 cluster members have been confirmed with a median redshift of
 $z=1.237$ and a velocity dispersion $\sigma_v\approx 800$ km/s.

\subsection{\Chandra data}

RDCS1252 was observed with the \Chandra ACIS--I detector in VFAINT
mode in two exposures of 26 ks (Obs ID 4403) and 163 ks (Obs ID 4198).
The data were reduced using the the CIAO software V2.3 (see {\tt
http://cxc.harvard.edu/ciao/}) starting from the level 1 event file.
We used the tool {\tt acis\_process\_events} with the {\tt vfaint=yes}
option to flag and remove bad X-ray events which are mostly due to
cosmic rays.  Such a procedure reduces the ACIS particle background
significantly compared to the standard grade selection\footnote{see
{\tt http://asc.harvard.edu/cal/Links/Acis/acis/Cal\_prods/vfbkgrnd/}},
whereas source X-ray photons are practically unaffected (only
$\sim\!2\%$ of them are rejected, independently of the energy band,
provided there is no pileup).  We also applied the correction to the
charge transfer inefficiency to partially recover the original spectral
resolution of ACIS--I.
The data were filtered to include only the standard event grades 0, 2,
3, 4 and 6.  We removed $\sim 10$ hot columns via visual inspection.
We searched for flickering pixels (defined as those with more than two
events contiguous in time, where a single time interval was set to 3.3
s), however most of them are already removed by the filtering of bad
events for exposures taken in VFAINT mode.  We then applied a
3--$\sigma$ clipping filtering of time intervals with high background
levels using the script {\tt analyze\_ltcrv}, part of the CIAO
distribution.
The total effective exposure time is 188 ks after the application of
this reduction procedure.

In Fig.~\ref{f:xfield}, we show part of the ACIS-I field.  The diffuse
X-ray emission from the cluster is detected with a high
signal-to-noise (S/N peaks to $\sim 21$ within a radius of 35\arcsec),
and can be traced out to $r=59\arcsec$ ($2\sigma$ above the
background).  The \Chandra image (see bottom left of
Fig.~\ref{f:xfield}) immediately shows that point sources do not
significantly contaminate the cluster emission, as it is sometimes the
case (e.g. Stanford et al. 01). In most cases, sources close to the 
cluster core in projected distance have been found to be
foreground or background AGN. In the RDCS1252 field we have only a
few identifications of point sources to date (8 within a
$6\arcmin\times 6\arcmin\,$ area), of which only one is at the cluster
redshift (Fig.~\ref{f:xfield}).

We performed the spectral analysis in two circular regions (35\arcsec\
and 59\arcsec\ radii) around the centroid of the photon distribution
after masking out point sources. In these apertures we detected
approximately 850 and 1220 net counts in the 0.3--10 keV band.  The
background is obtained from a large annulus around the cluster
position, after subtraction of point sources.  The background photon
file is scaled to the source file by the ratio of the geometrical
area.  We checked that variations of the background intensity across
the chip do not affect the background subtraction, by comparing the
count rate in the source and in the background at energies larger than
8 keV, where the signal from the source is null.  The response
matrices and the ancillary response matrices were computed for each
exposure with the tool {\sl acisspec} applied to the extraction
regions.  We applied the script {\tt apply\_acisabs} by Chartas and
Getman to take into account the degradation in the ACIS QE due to
material accumulated on the ACIS optical blocking filter since
launch\footnote{See {\tt
http://cxc.harvard.edu/ciao/threads/apply\_acisabs/}}.  We manually
decreased the effective area below 1.8 keV by 7\% to homogenize the
low--energy calibrations of ACIS--S3 and ACIS--I (see Markevitch \&
Vikhlinin 2001).

\subsection{\XMM data}
The XMM--Newton observations were carried out in two epochs, on
January 1 and 11 2003, for a total of $69.71+69.71=139.42$ ksec, using
the European Photon Imaging Camera (EPIC) PN and MOS detectors
(observation Id 0057740301/401). The PN data were negatively affected
by one of the CCD gaps which fell on the outskirt of RDCS1252, making
it difficult to extract regions for a merged MOS+PN spectroscopic
analysis. After experimenting with different apertures, masking and
background subtraction techniques, we decided to use the MOS1$+$MOS2
only for this analysis to avoid systematics which are at present not
fully understood. Nonetheless, as shown below, data from the two MOS
detectors significantly improved the signal-to-noise of the extracted
spectra when combined with the \Chandra data.  We used the XMM
Standard Analysis System (SAS) routines (SASv5.4.1) to obtain
calibrated event files for the MOS1, MOS2  cameras.  Time
intervals in which the background was increased by soft proton flares
were excluded by rejecting all events whenever the count rate exceeded
20 cts/100s in the 10--12 keV band for each of the two MOS cameras.  
The final effective exposure time amounts to
137 ks for the two MOS detectors.  The spectrum
was extracted from an aperture of 42\arcsec\ radius (see
Fig.~\ref{f:xfield}), which avoids all point sources clearly visible
in the \Chandra image.  In this aperture, we measured 2110 MOS1$+$MOS2
net counts in the 0.5-8 keV band used for spectral fitting (1410 net
counts in 0.5-2 keV band). Comparison of the \Chandra and \XMM images
shows the well known complementarity of the two observatories: \XMM
has lower sensitivity to point sources and is prone to confusion,
however its large collecting area yields high count rates on extended
sources, much needed for spectroscopic analysis. \Chandra allows ICM
morphology and profiles to be studied even at these large redshifts,
by separating the diffuse component from faint field sources.

\section{RESULTS}

\subsection{Spectral analysis}

The spectra are analyzed with XSPEC v11.2.0 (Arnaud 1996) and fitted
with a single temperature MEKAL model (Kaastra 1992; Liedahl et
al. 1995), where the ratio between the elements are fixed to the solar
value as in Anders \& Grevesse (1989).  These values for the solar
metallicity have recently been superseded by the new values of
Grevesse \& Sauval (1998), who used a 0.676 times lower Fe solar
abundance.  However, we prefer to report metallicities in units of the
Anders \& Grevesse abundances since most of the literature still
refers to these old values.  Since our metallicity depends only on the
Fe abundance, updated metallicities can be obtained simply by
rescaling by 1/0.676 the values reported in Table 1.  We model the
Galactic absorption with the tool {\tt tbabs} (see Wilms, Allen \& McCray
2000).

The fits are performed over the energy range 0.6--8 keV.  We exclude
photons with energy below 0.6 keV in order to avoid systematic biases
in the temperature determination due to uncertainties in the ACIS
calibration at low energies.  We used three free parameters in our
spectral fits: temperature, metallicity and normalization.  We freeze
the local absorption to the Galactic neutral hydrogen column density
$N_H = 5.95 \times 10^{20}$ cm$^{-2}$, as obtained from radio data
(Dickey \& Lockman 1990), and the redshift to $z=1.237$, as measured
from the optical spectroscopy.  Spectral fits are performed using the
Cash statistics (as implemented in XSPEC) of source plus background
photons, which is preferable for low signal--to--noise spectra.
  We also performed the same fits with the $\chi^2$ statistics (with a
standard binning with a minimum of 20 photons per energy channel in
the source plus background spectrum) and verified that our best--fit
model always gives a reduced $\chi^2 \sim 1$. All quoted errors below
correspond to $1\sigma$, or 68\% confidence level for one interesting 
parameter.

The \Chandra folded and unfolded spectra of RDCS1252 are shown in
Fig.~\ref{f:xspec} for the larger aperture. A prominent Fe K is
visible at $kT\simeq 3$ keV, which represents the first clear
detection of an iron line from an ICM at $z>1$.  Using only \Chandra
data, the fit to the spectrum in the inner 35\arcsec\ radius gives a
best fit temperature of $kT = 6.4^{+1.0}_{-0.8}$ keV, and a best fit
metallicity of $Z = 0.47^{+0.21}_{-0.18} \, Z_\odot$.  
Our Fe K-line diagnostic is simpler and more robust
than that based on the line--rich region around 1 keV, where the line
emission is dominated by the L--shell transition of Fe, and the
K--shell transitions of O, Mg, and Si.  As a consistency check, if we
leave free the Galactic absorption, we obtain a best fit value of $N_H
= 2.9 \times 10^{20}$ cm$^{-2}$, with an upper limit of $ 6.3 \times
10^{20}$ cm$^{-2}$ at $1\sigma$, thus consistent with the Galactic
value of $N_H = 5.95 \times 10^{20}$ cm$^{-2}$.  In this case, the
best fit temperature is consistent (within $1\sigma$) with the
aforementioned best fit value. Interestingly, leaving the redshift
free, a four parameter fit yields $z = 1.234_{-0.035}^{+0.033} $,
which shows how accurately the redshift can be determined from the
X-ray data alone due to the high signal-to-noise detection of the Fe
line.

It is useful to repeat the analysis using the larger aperture to test
systematic effects in the background subtraction. The fit to the
spectrum extracted from the 59\arcsec\ aperture gives a temperature
somewhat lower, $kT = 5.2^{+0.7}_{-0.6}$ keV, and a metallicity of $Z
= 0.64^{+0.20}_{-0.18} \, Z_\odot$.  The difference between the
temperature measurement in the two apertures is not significant enough
to be attributed to a temperature gradient. We also measured
temperatures in two independent radial bins ($r<35\arcsec,\,
35\arcsec<r<59\arcsec$) and could find only weak evidence ($1\sigma$)
of a temperature drop outward. The best fit redshift in this case is
still consistent within $1\sigma$ with the spectroscopic redshift:
$z=1.27_{-0.03}^{+0.02}$.  The spectral analysis of the \Chandra data
also yields a flux within the 59\arcsec\ aperture of $2.9\times
10^{-14}\fun $ in the 0.5-2 keV band, in good agreement with the ROSAT
value. This corresponds to a luminosity of $1.9\times 10^{44}\lun
h_{70}^{-2}$ in the rest frame 0.5-2 keV band, and a bolometric
luminosity of $6.6\times 10^{44}\lun h_{70}^{-2}$. One can use the
best fit $\beta$-model of the surface brightness profile described
below to extrapolate these luminosities at larger radii. For example,
values need to be multiplied by a factor 1.3 to encircle the flux
within $r=1$ Mpc (or 2\arcmin).

Using the \XMM data from the MOS detectors only, we extracted a
spectrum in the energy range 0.5-8 keV (containing 2110 counts as
opposed to 800 in the \Chandra spectrum) and verified that
the best fit temperature and metallicity are consistent, within
$1\sigma$, with the \Chandra mesurements above. 
The XMM/MOS spectrum is shown in Fig.~\ref{f:xspec_xmm}.
 To enhance the photon
statistics, we performed a combined fit of the \Chandra spectrum
extracted from the 59\arcsec\ region, and the two MOS spectra from the
137 ksec \XMM observations.  Thus, we obtain a best fit temperature $kT =
6.0^{+0.7}_{-0.5}$ keV, and a best fit metallicity $Z =
0.36^{+0.12}_{-0.10} \, Z_\odot$.  If we leave the redshift free, we
obtain a best fit value of $z = 1.221^{+0.024}_{-0.017}$, i.e.
within 1\% of the spectroscopic redshift.

In Fig.~\ref{f:contours}, we show confidence contours in the $Z-kT$,
$z-kT$ planes relative to the spectral fits discussed above. The
combined \Chandra and \XMM analysis yields a temperature accuracy of
10\%, which is unprecedented at these redshifts. 
  We defer the analysis of extended and point sources from
the full \XMM data set, combined with \Chandra and HST-ACS
observations of the field (Blakeslee et al. 2003), to another paper
(Mainieri et al. in preparation).

For completeness, we briefly report on the serendipitous group
CXJ1252.6\minus2924 located at $12^h52^m34.2^s \; \minus29^\circ
24\arcmin 59\arcsec$ (J2000), 2.5\arcmin\ NW of RDCS1252 (see
Fig.~\ref{f:xfield}).  Extracting a spectrum from the \Chandra data
with an aperture of 30\arcsec\ radius, we detect several low
ionization metal lines. The best fit MEKAL model yields: $kT =
1.6^{+0.16}_{-0.31}$ keV, $Z = 0.42^{+0.33}_{-0.35}\, Z_\odot$, and a
redshift $z = 0.32^{+0.02}_{-0.10}$. In this aperture, we measure 320
net counts, a flux of $8.7\times 10^{-15}\fun $ in the 0.5-2 keV band,
corresponding to $L_X[0.5-2\, {\rm keV}]=3.5\times 10^{42}\lun$, and
$L_{\rm bol}=6.3\times 10^{42}\lun$. Inspection of our CTIO I band
image shows a galaxy group dominated by a luminous elliptical galaxy
at the X-ray centroid.

\subsection{Mass determination}

The angular resolution provided by \Chandra and the detection of the
X-ray emission out to 1\arcmin\ (i.e. 0.5 Mpc) radius allows an
accurate modeling of the gas profile of RDCS1252.  This information,
assuming hydrostatic equilibrium and isothermality of the gas, leads
to a robust estimate of the total mass. 
The surface brightness profile is obtained from the exposure-corrected
image by fixing the number counts per bin to 50 and is fitted with an
isothermal $\beta-$model (Cavaliere \& Fusco-Femiano 1976) providing a
core radius of $79 (\pm 13) h_{70}^{-1}$ kpc (or 9.5\arcsec\ ) and
$\beta=0.529 (\pm0.035)$ (see Fig.~\ref{f:sb} and Ettori et al. 2003b
for details).  The model provides a reasonably good fit to the profile
($\chi^2=168$, with 156 degree of freedom), which does not require any
further component (e.g. a double $\beta-$model) from a statistical
point of view.  We recover the gas density and total gravitating mass
profile using analytic formula associated with the
$\beta-$model: $n_{\rm gas} = n_{0, \rm gas} (1+x^2)^{-3 \beta/2}$ and
$M_{\rm tot} = \frac{3 \ \beta \ T_{\rm gas} \ r_{\rm c}}{G \mu m_{\rm
p}} \frac{x^3}{ 1+x^2 }$, where $x=r/r_{\rm c}$, $\mu$ is the mean
molecular weight in atomic mass unit ($=0.6$), $G$ is the
gravitational constant, $m_{\rm p}$ is the proton mass; the central
density, $n_{0, \rm gas}$, is obtained from the combination of the
best-fit results from the spectral and imaging analyses as described
in Ettori, Tozzi \& Rosati (2003a). The errors
are obtained from the distribution of the values after 1\,000
Monte-Carlo simulations.  We measure masses within the radius
$R_\Delta$ encompassing a fixed density contrast, $\Delta_z=500$ in an
Einstein-de Sitter universe, with respect to the critical density,
$\rho_{c,z}$, i.e.  $\Delta_z = 3M_{\rm
tot}(<R_\Delta)/(4\pi\rho_{c,z} R_\Delta^3)$.  At $R_{500}=536\pm 40$
kpc, corresponding to an overdensity of 457 at the cluster redshift
(or $\Delta=500$ in an Einstein-de Sitter universe), we measure
$M_{\rm gas} = (1.8 \pm 0.3) 10^{13} h_{70}^{-5/2} M_{\odot}$ and
$M_{\rm 500} = (1.6 \pm 0.4) 10^{14} h_{70}^{-1} M_{\odot}$. 
These values are associated with the
\Chandra temperature measurement, $T_{\rm gas}=5.2$ keV.
If we
take the best fit temperature of 6.0 keV from the combined \Chandra
and \XMM analysis, the total mass scales up accordingly and the error
bar decreases:
$M_{500}=(1.9\pm 0.3) 10^{14} h_{70}^{-1} M_{\odot}$.  In order to
estimate the cluster virial mass, we can extrapolate our mass
measurement to larger radii using a typical Navarro, Frenk \& White
profile with concentration $c=5$. This yields $M_{\rm vir}\approx
M_{\Delta=200}=1.4\, M_{\rm 500}=2.7\times 10^{14} h_{70}^{-1}
M_{\odot}$.  We also find a gas mass fraction $f_{\rm gas}=(0.10\pm
0.04) h_{70}^{-3/2}$, consistent with other measurements in distant
clusters (e.g. Ettori et al. 2003a).

\section{DISCUSSION AND CONCLUSIONS}

Taking advantage of the complementarity between \Chandra and \XMM
observations, we have measured physical properties of RDCS1252 at
$z=1.237$ with unprecedented accuracy at these redshifts. The \Chandra
data allow the gas profile to be traced and modeled out to $\sim\!
500$ kpc, free of confusion from field sources, and therefore enables
the mass to be accurately derived. By augmenting our spectroscopic
analysis with the \XMM data (MOS detectors only), we narrowed down the
$1\sigma$ error bar to 10\% for the temperature and 30\% for the
metallicity.  In Table 1, we report
a summary of the main physical properties of RDCS1252 measured from
the X-ray data.

\begin{figure}
  \begin{center}
\includegraphics[width=0.7\columnwidth,angle=0]{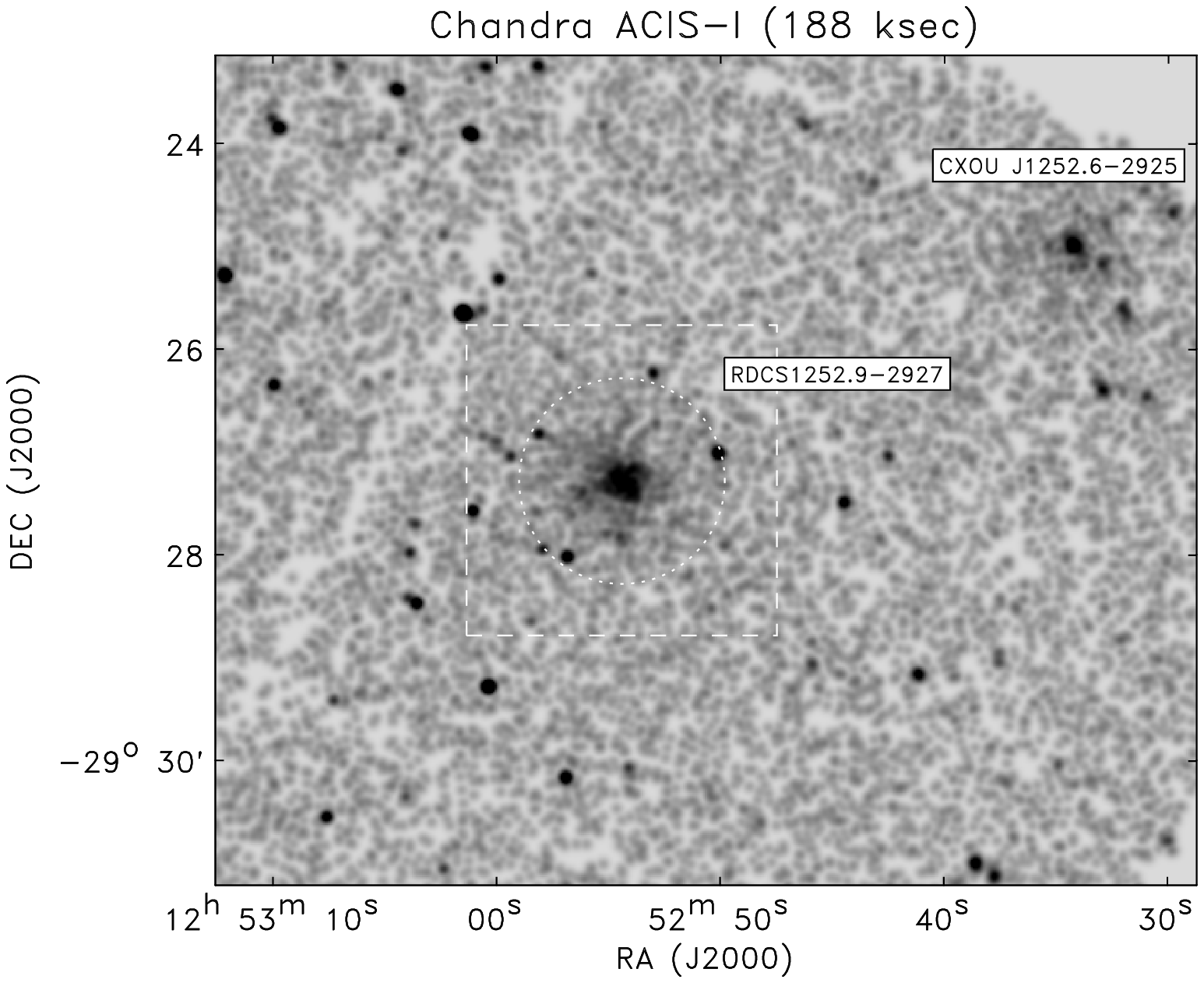}\\
\hspace*{7mm}\hbox{
    \includegraphics[width=0.4\columnwidth,angle=0]{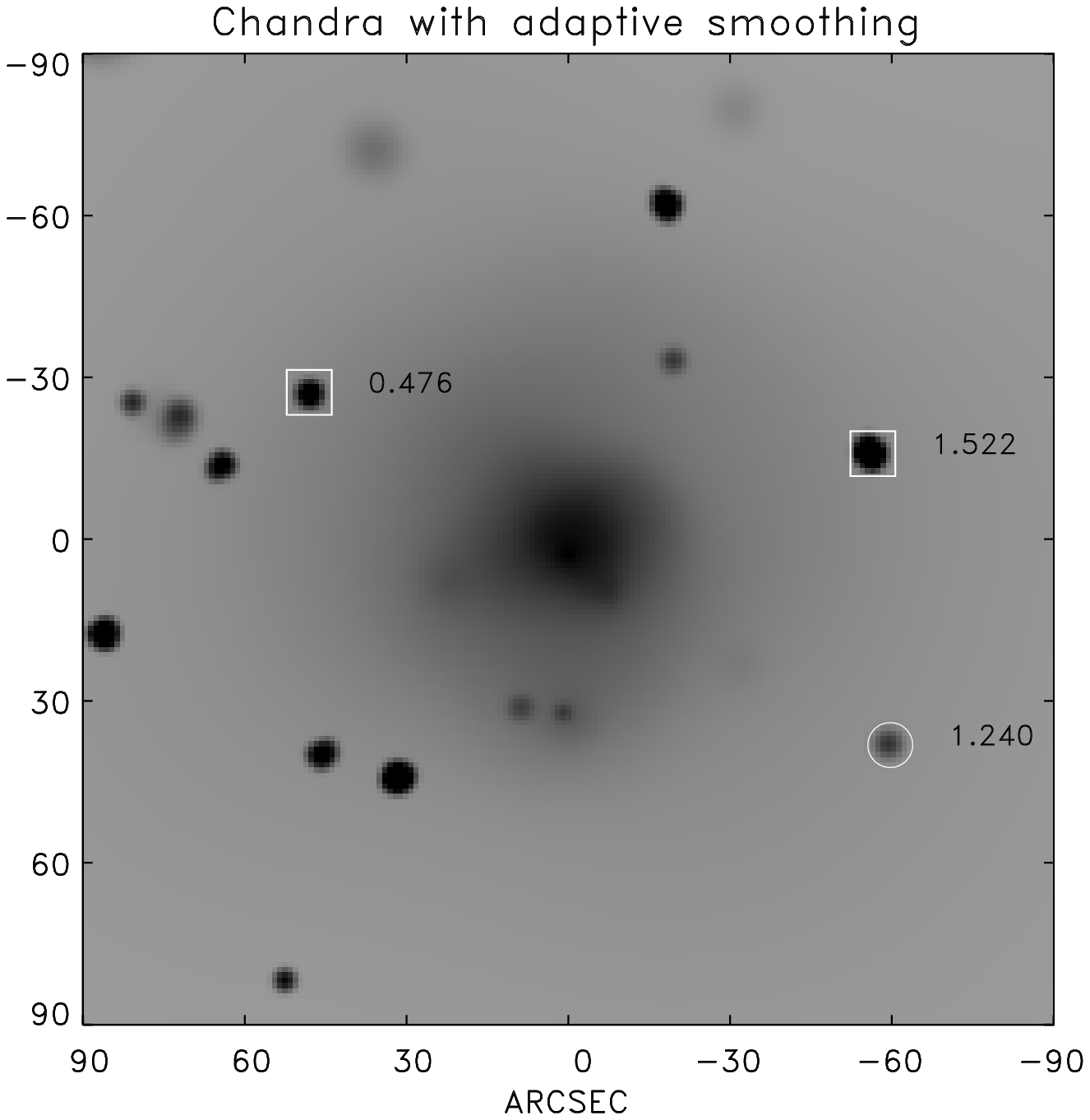}\hfil
    \includegraphics[width=0.4\columnwidth,angle=0]{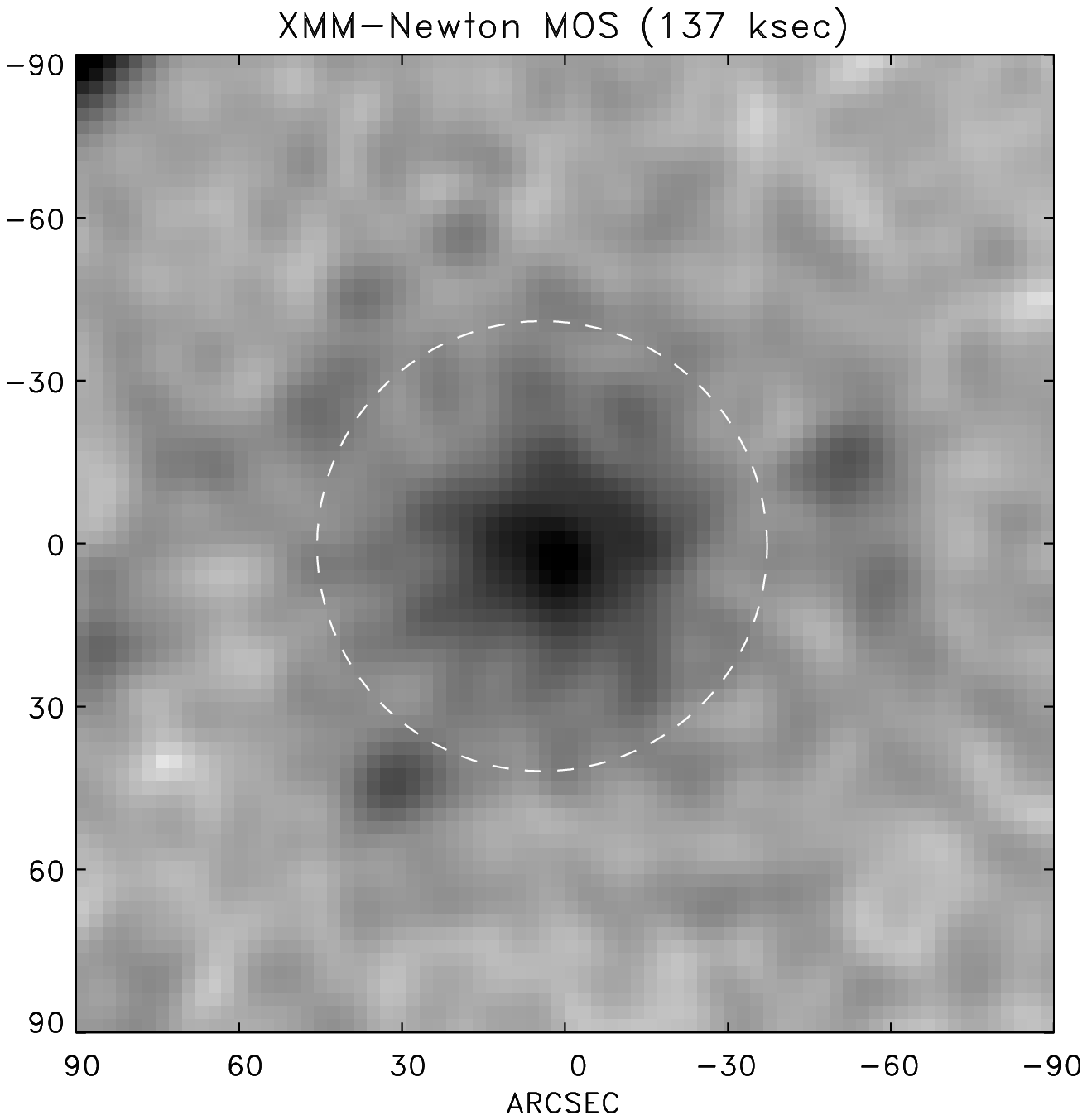} }
  \end{center}
\caption{Top: Grey scale image of
\Chandra ACIS-I 188 ksec observations in the 0.5-2 keV band showing
$9.6\arcmin\times 8.1\arcmin$ field around RDCS1252.9\minus2927 (a
serendipitous low redshift group, CXOU J1252.6\minus2925, is also
visible). The image has been smoothed by a gaussian with
$\sigma=2\arcsec$ and the grey-scale has a square root scaling; the
dashed circle shows the $59\arcsec$ aperture used to extract
spectra. The box marks a $3\arcmin\times 3\arcmin$ area around
RDCS1252 shown in the bottom panels. Left: adaptively smoothed
\Chandra image in logarithmic scale. Sources with spectroscopic
redshift are marked, the faint one (circle) is at the cluster
redshift.  Right: \XMM MOS image (137 ksec) in the 0.5-2 keV band.
The image has been smoothed by a gaussian with $FWHM=9\arcsec$ and the
grey-scale has square root scaling; the spectroscopic aperture of
42\arcsec\ radius is also shown. }
\label{f:xfield}
\end{figure}

\begin{figure}
     \centering 
    \includegraphics[width=0.5\columnwidth,angle=0]{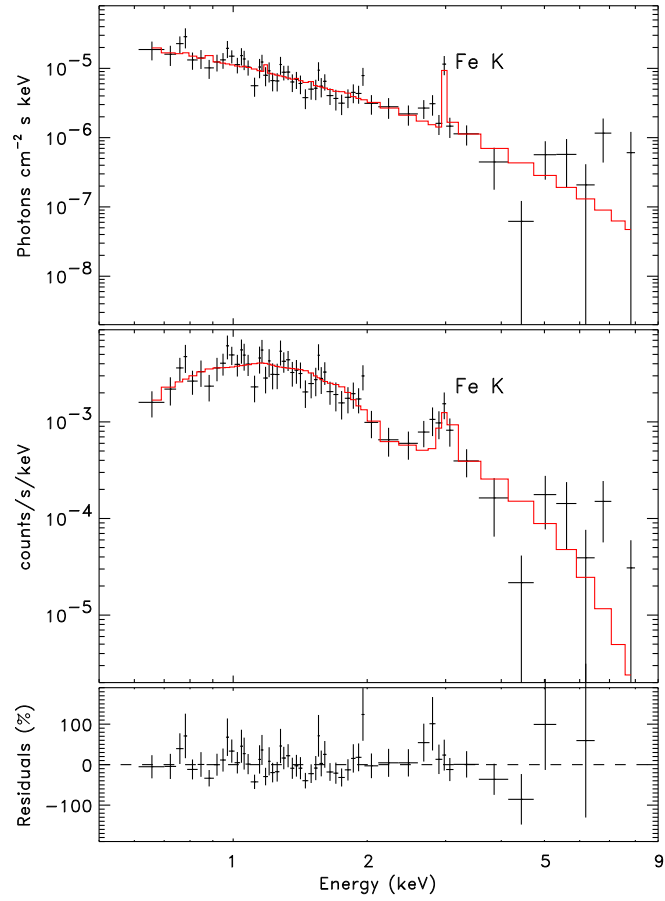}\
\vspace*{-1.5cm}
\caption{X-ray spectrum (data points) and best fit MEKAL model
(solid line) from \Chandra observations (188 ksec) of
RDCS1252\minus2927 at $z_{\rm spec}=1.237$; from top to bottom:
unfolded, folded spectrum and relative residuals; A clear redshifted Fe
6.7 keV line is visible. The spectrum is extracted from a 59\arcsec\ radius
region. }
\label{f:xspec}
\end{figure}

\begin{figure}
     \centering 
    \includegraphics[width=0.5\columnwidth,angle=0]{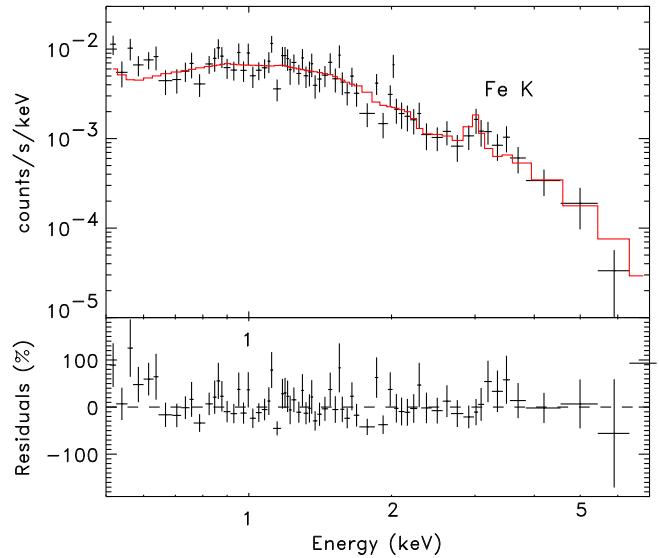}\
\vspace*{-1.5cm}
\caption{X-ray spectrum (data points) and best fit MEKAL model
(solid line) from \XMM observations (MOS detectors only, 137 ksec) of
RDCS1252\minus2927. The spectrum is extracted from a 42\arcsec\ radius
region. }
\label{f:xspec_xmm}
\end{figure}

\begin{figure}
  \centering\hbox{
  \includegraphics[width=.4\columnwidth,angle=0]{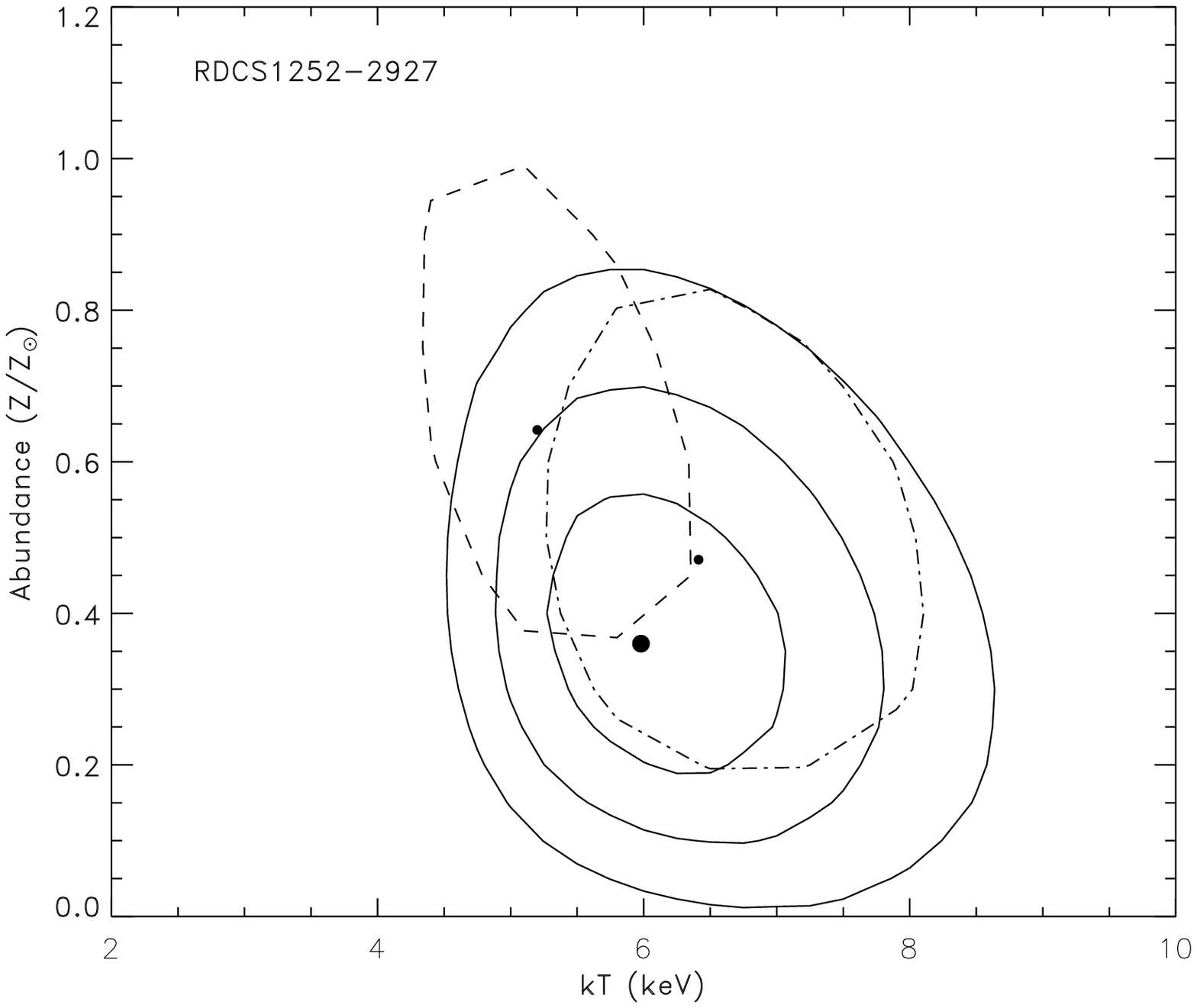} \hfil
  \includegraphics[width=.4\columnwidth,angle=0]{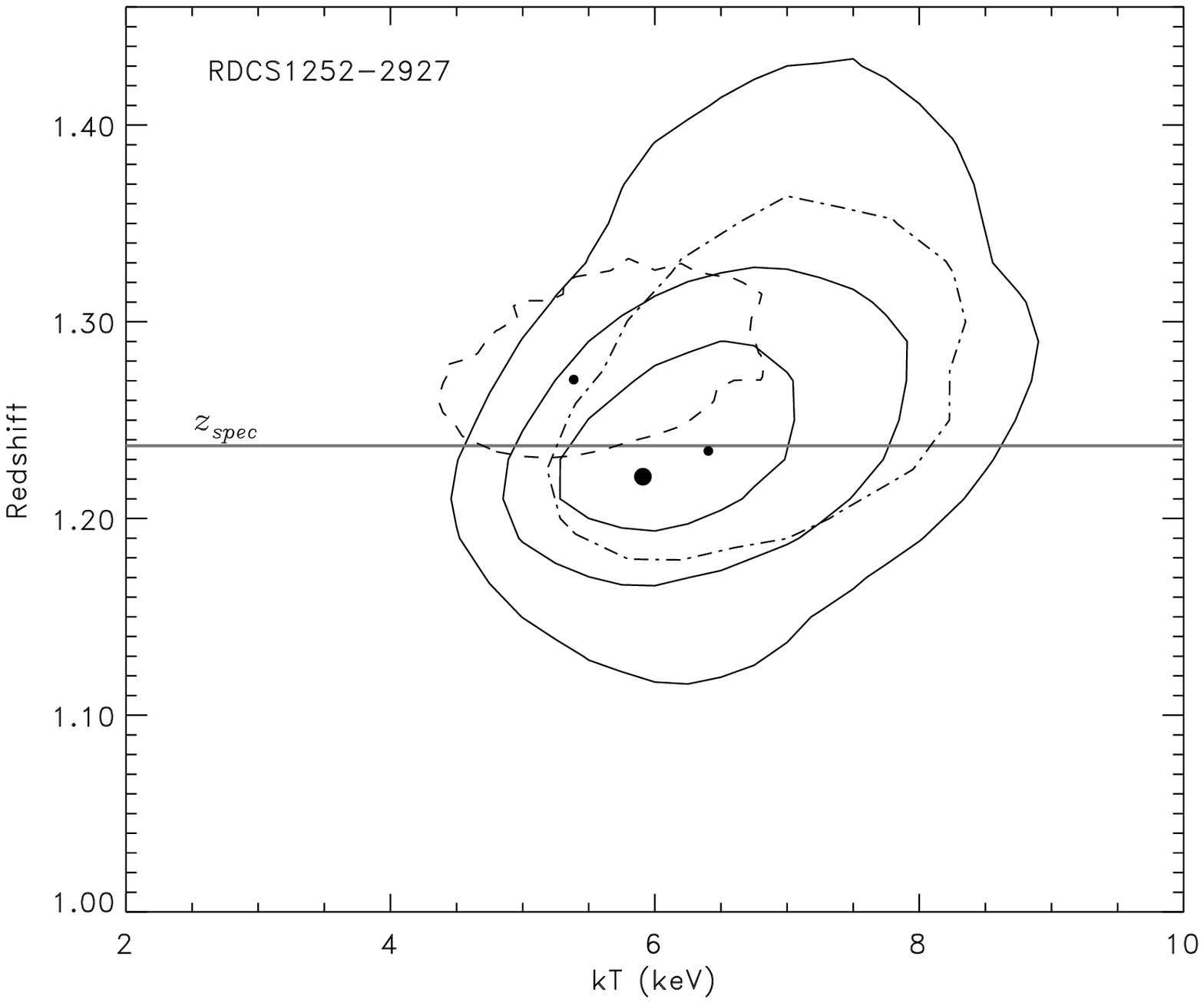} }
\caption{Left: best fit temperature and
 metallicity of the gas obtained by combining \Chandra ACIS-I and {\it
 XMM-Newton} MOS data (solid contours for $1, 2, 3\sigma$ confidence
 levels for two interesting parameters). Dashed (dot-dashed) contours
 show the 1$\sigma$ levels obtained from the \Chandra data only, with
 apertures of 60\arcsec\ (35\arcsec) radius. Right: best fit redshifts
 and temperatures of the ICM, with the horizontal line marking the
 spectroscopic redshift based on 36 cluster members. }
\label{f:contours}
\end{figure}

\begin{figure}
 \centering
\hbox{\hspace*{-1cm}
\includegraphics[width=0.5\columnwidth,angle=0]{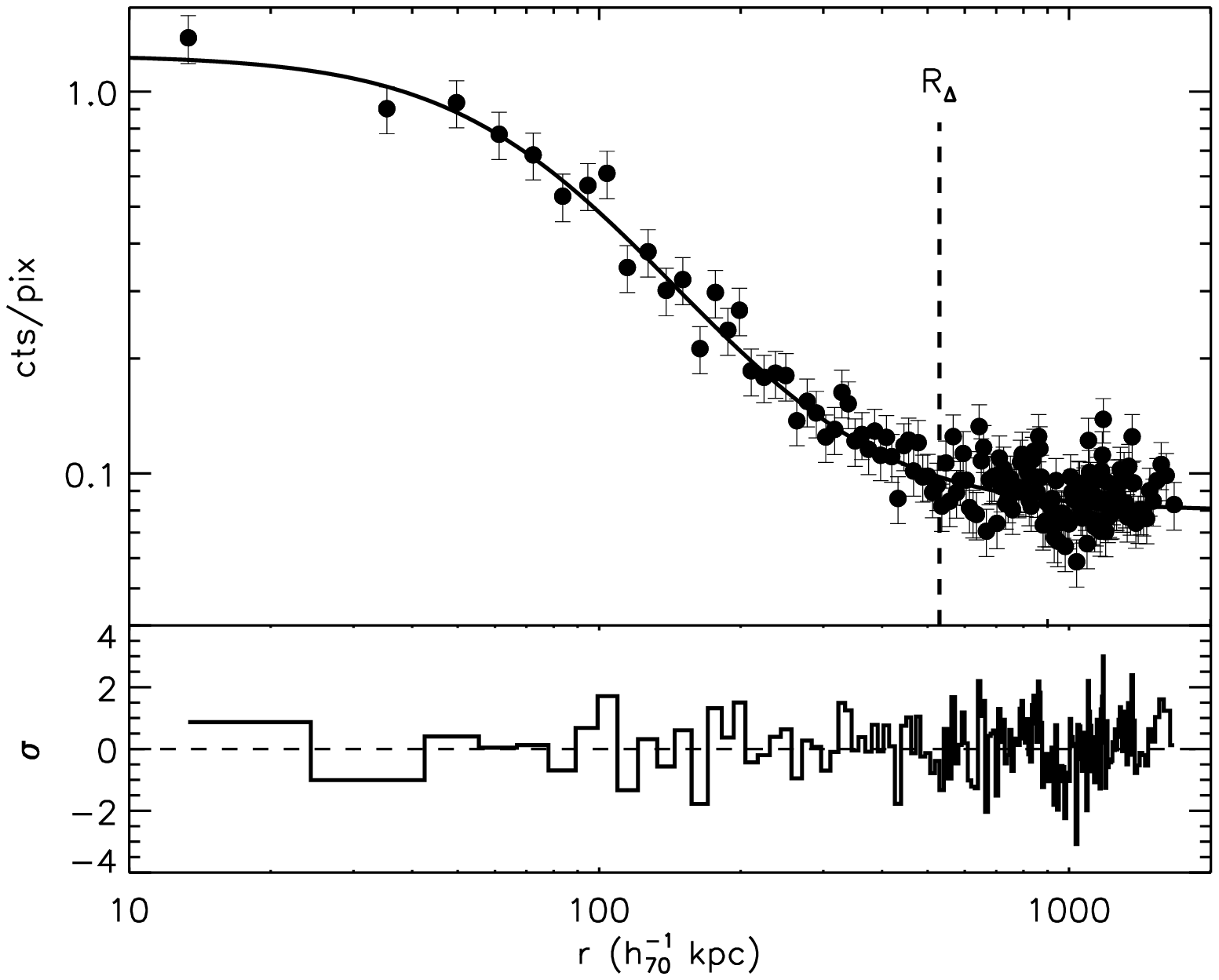}
\includegraphics[width=0.4\columnwidth,angle=0]{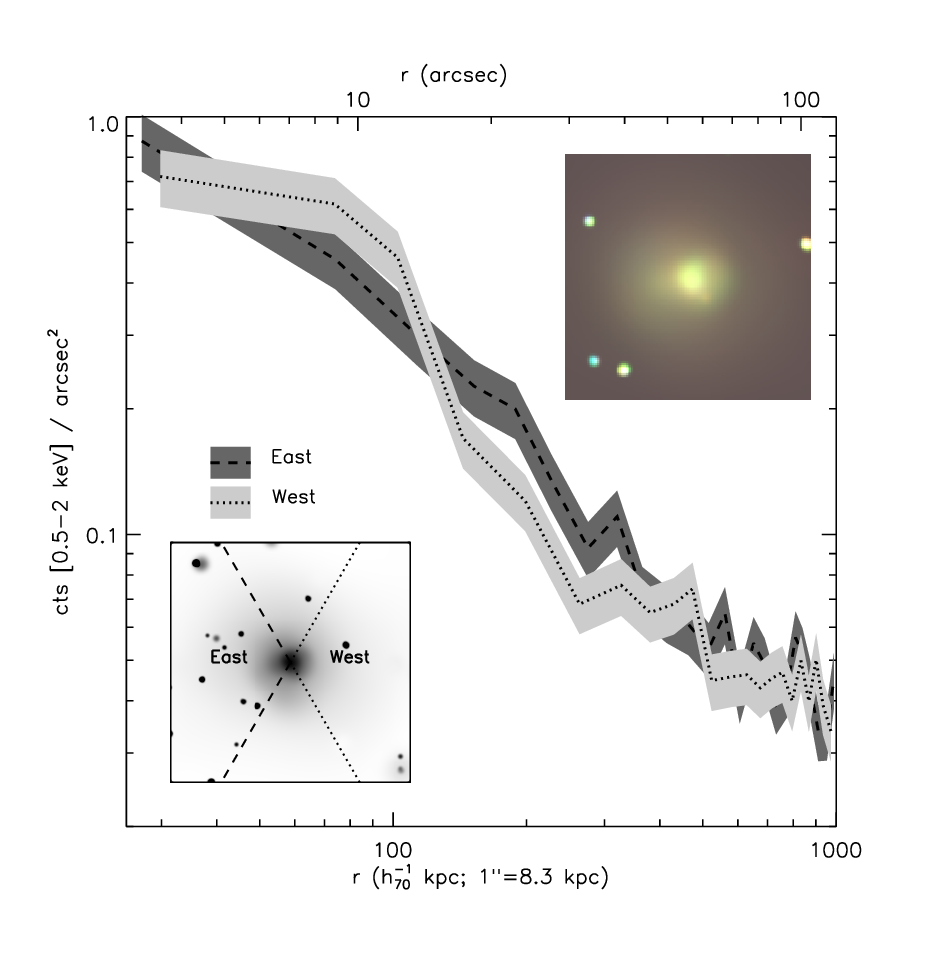} 
}
\caption{Left panel: surface brightness profile of
RDCS1252.9\minus2927 with best fit $\beta$-model (solid line) and
residuals. $R_\Delta=R_{500}$ indicates the radius within which the
total and gas mass are calculated (1\arcmin\ corresponds to 500 kpc at
$z=1.24$ for the adopted cosmology). Right panel: surface brightness
profiles azimuthally averaged in two separate sectors over the area
shown in the lower left inset; the shaded areas correspond to
$1\sigma$ error bar. The upper right inset shows the adaptively
smoothed, X-ray color image of the cluster core ($2\arcmin\times
2\arcmin$) showing the asymetric distribution of X-ray emission (see
text). }
\label{f:sb}
\end{figure}

In Fig.~\ref{f:overlay}, we show a color composite image of the field
with overlaid \Chandra contours. The color image combines deep
near-infrared imaging with ISAAC and optical imaging with FORS at the
VLT, with limiting AB magnitude of $\sim\! 26$ (Lidman et al. 2003,
Rosati et al. in preparation). Cluster early type galaxies stand out
as red objects which cluster strongly toward the centroid of the X-ray
emission. The two central galaxies, which are 2\arcsec\ apart with
Vega magnitudes $K\simeq 17.5$, lie near the peak of the X-ray
emission. The overall distribution of cluster galaxies, well mapped by
our spectroscopic and photometric redshifts, appear flattened along
the E-W direction (Toft et al. in prep.). A close inspection of the
\Chandra data reveals an interesting feature in the X-ray surface
brightness distribution of RDCS1252. In Fig.~\ref{f:sb}, the surface
brightness profile azimuthally averaged in two separate sectors shows
a discontinuity on the west side, at $r\approx 15\arcsec$ (or 125
kpc). This could be the origin of the relatively low $\beta$ value
(0.53) obtained by the King profile fit. A relatively sharp edge on
the west side is visible in the raw \Chandra image
(Fig.~\ref{f:xfield}), and is apparent in the adaptively smoothed
X-ray color image (see upper right inset of Fig.~\ref{f:sb}), which is
the composite of the three energy bands [0.5-1], [1-2], [2-7] keV.  A
close inspection of this image reveals a comet-like shape of the X-ray
emission in the cluster core, a feature resembling the remarkable
shock front discovered in the cluster 1E0657--56 at $z=0.3$
(Markevitch et al. 2002), which is the result of a merging process of
a cluster subclump. However, the relatively low photon statistics of
the \Chandra data prevent us from further speculating on the physical
nature of this feature in RDCS1252. No major subclumps are visible in
the distribution of the cluster galaxies, however we note the
coincidence between the mild E-W asymetry of the gas and the E-W
elongation of the cluster members. This could be the result of the
infall of cluster galaxies along a major filament associated with a
cold-front morphology of the gas due to the merger of a sub-clump just
exiting the cluster core along the E-W direction.

The physical properties of RDCS1252, as derived from the \Chandra
data, help to constrain the high redshift end of cluster scaling
relations and study their evolution. For example the $L_X-T$, $M-T$
relations, the entropy and metallicity of the ICM as function of
redshift. We refer to the analysis of Ettori et al. (2003b), which
used a sample of 26 clusters at $z>0.4$, also including RDCS1252. We
note here that the values of $L_{\rm bol}$ and $T$ of RDCS1252 are
consistent with the local $L-T$ relation measured by Markevitch
(1998), and when combined with all the other data available on distant
clusters in the \Chandra archive, suggest only a mild positive
evolution of the $L_X-T$ relation (see discussion in Ettori et
al. 2003b and references therein). Our best fit value of the
metallicity, $Z =0.36^{+0.12}_{-0.10} \, Z_\odot$, lends further
support to a lack of evolution of the ICM mean metallicity measured
out to $z\simeq 1.3$ (see analysis by Tozzi et al. (2003), which did
not include RDCS1252).

Overall, our analysis implies that RDCS1252 is the most X-ray luminous
and likely the most massive bona-fide cluster discovered to date at
$z>1$. Despite the large look-back times probed by these observations,
RDCS1252 appears already well thermalized, with thermodynamical
properties, as well as metallicity, very similar to those of clusters
of the same mass at low redshift. This is consistent with a scenario
in which the major episode of metal enrichment and gas preheating by
supernova explosions occurred at $z\sim\! 3$.

X-ray selected cluster surveys in the ROSAT era have led to routine
identification of clusters out to $z\simeq 0.85$, with only a few
examples at higher redshifts (Rosati et al. 1999, Ebeling et al. 2001,
Stanford et al. 2002, Rosati et al. 2002).  Although the redshift
boundary for X-ray clusters has receded to $z=1.3$ recently, a census
of clusters at $z\simeq 1$ has just begun and the search for clusters
at $z>1.3$ remains a serious observational challenge. Extrapolating
the RDCS yield to \XMM or \Chandra based serendipitous surveys now
underway (e.g. Romer et al. 2001, Boschin 2002), one expects $\sim\!
10$ clusters as luminous as RDCS1252 in a 50 deg$^2$ area.  An
inspection of the underlying cluster mass function at $z=1.24$
(e.g. Borgani et al. 2001), for our adopted cosmology and
$\sigma_8=0.7-0.8$, shows that RDCS1252, with $M_{\rm vir}\sim\!
3\times 10^{14} M_{\odot}$, could well represent a typical $M^\ast$
cluster at these redshifts. Moreover, we note that the presence of
such a cluster in the RDCS survey volume (see Fig.5 in Rosati et
al. 2002) is in agreement with predictions based on the current
cosmological concordance model (e.g. Bennett et
al. 2003). Specifically, for the assumed cosmology and $\sigma_8=0.8$,
we expect to find one cluster as massive as RDCS1252 or
more in $1.5\times 10^7 (h_{70}^{-1}{\rm Mpc})^3$ at $z\simeq 1.2$.

 Using high-$z$ radio galaxies as signposts for proto-clusters has
been the only viable method so far to break this redshift barrier and
push it out to $z\simeq 4$ (e.g., Venemans et al. 2002, Kurk et
al. 2003).  If there is an evolutionary link between these strong
galaxy overdensities around distant radio galaxies and the X-ray
clusters at $z\simeq 1.2$, viable evolutionary tracks should be found
linking the galaxy populations in these systems, using their
spectrophotometric and morphological properties.  Recent follow-up
\Chandra observations of high redshift radio galaxies have revealed
the presence of diffuse X-ray emission, in addition to a central point
source (3C294 at $z=1.786$: Fabian et al. 2003a; 4C41.17 at $z=3.8$:
Scharf et al. 2003). However, their spectral energy distribution and
other energetic arguments indicate that the extended emission is
likely non-thermal, and due instead to inverse Compton scattering of the
CMB photons by a population of relativistic electrons associated with
the radio source activity. The serendipitous detection of thermal ICM
at $z> 1.5$ associated with $\sim\!L^\ast$ clusters remains extremely
difficult, not only for the lack of volume in current X-ray surveys,
but also for the severe $(1+z)^4$ surface brightness dimming which
affects X-ray observations. These limitations will eventually be
overcome by surveys exploiting the Sunyaev--Zeldovich (SZ) effect
(e.g. Carlstrom et al. 2002), which will explore large volumes at
$z>1$. It is worth noting, however, that the current sensitivity of SZ
observations is still not sufficient to detect any of the known X-ray
clusters at $z>1$, all having $L_X[0.5-2\,{\rm keV}]\lesssim 3\times
10^{44}\lun$, such as RDCS1252.  The current generation of large area
optical surveys (e.g. using the $z$-band; Gladders \& Yee 2000) remain
a valid alternative to unveil a sizeble number of clusters at $z\sim\!
1$, while the next generation of large area surveys in the near-IR 
(e.g. Warren 2002) will
push this boundary even further. Without a correspondingly large area
X-ray survey, however, our ability to glean physical properties
necessary to test structure formation scenarious, as well deriving
cosmological parameters, will be rather limited.

\acknowledgements The authors aknowledge support from NASA grant
GO3-4166A. PR thanks the CXC for their prompt help in planning the
\Chandra observations. 
Part of this work was performed
under the auspices of the U.S.\  Department of Energy by University of
California, Lawrence Livermore National Laboratory under contract
No.\ W-7405-Eng-48.
CN acknowledges support under the ESO
visitor program in Garching during the completion of this work.
RDC acknowledges partial financial support from
ASI (I/R/062/02).

\begin{deluxetable}{lcccccc}
\tablewidth{0pt} 
\tablecaption{X-ray Properties of RDCS1252.9\minus2927 at $z=1.237$ \tablenotemark{\ast}} 
\tablehead{ 
\colhead{RA \qquad\qquad Dec} & 
\colhead{$L_{[0.5-2.0]}$\tablenotemark{a}} & 
\colhead{$L_{[bol]}$\tablenotemark{a}} & 
\colhead{$T_x$ \tablenotemark{b}} & 
\colhead{$Z_{\rm gas}$\tablenotemark{b}} & 
\colhead{$M_{\rm gas}\tablenotemark{c}$} & 
\colhead{$M_{\rm tot}\tablenotemark{c}$} \\
\colhead{J2000} & 
\colhead{$10^{44}\lun$} &
\colhead{$10^{44}\lun$} & 
\colhead{keV} &
\colhead{$Z_\odot$} &
\colhead{$10^{13}M_{\odot}$} & 
\colhead{$10^{14}M_{\odot}$} 
} 
\startdata 
 $12^h52^m54.4^s \quad \minus29^\circ 27\arcmin 17\arcsec$
 & $1.9^{+0.3}_{-0.3}$ & $6.6^{+1.1}_{-1.1}$ 
 & $6.0^{+0.7}_{-0.5}$ & $0.36^{+0.12}_{-0.10}$
 & $1.8^{+0.3}_{-0.3}$ & $1.9^{+0.4}_{-0.4}$ 
\enddata
\label{tab1}
\tablenotetext{\ast}{Adopted cosmology: $H_0 = 70$ km s$^{-1}$ Mpc$^{-1}$,
  $\Omega_m=0.3$, and $\Omega_\Lambda=0.7$.}
\tablenotetext{a}{Luminosity within an aperture of 60\arcsec\ (or 500 kpc)}
\tablenotetext{b}{From combined \Chandra and \XMM spectral analysis}
\tablenotetext{c}{Mass measured out to $R_{500}=536\pm40$ kpc}
\end{deluxetable}

\newpage

\newpage

\newpage

\newpage

\newpage
 
\begin{figure}
 \centering
\fbox{{\bf See attached jpeg figure}}
\caption{Color composite image showing a $2\arcmin\times 2\arcmin$
field on RDCS1252.9\minus2927 at $z=1.24$ with overlaid \Chandra
contours. The image combines optical and near IR bands from the FORS
and ISAAC instruments on the VLT: $B+V$, $R+z$, and $J+K_s$. \Chandra
contours show the smoothed X-ray emission (with a gaussian FWHM of
5\arcsec) at the levels of 3, 5, 10, 20 $\sigma$ above the background. }
\label{f:overlay}
\end{figure}

\end{document}